\documentclass[conference]{IEEEtran}
\IEEEoverridecommandlockouts
\usepackage{cite}
\usepackage{amsmath,amssymb,amsfonts}
\usepackage{algorithmic}
\usepackage{graphicx}
\usepackage{textcomp}
\usepackage{xcolor}

\ifCLASSOPTIONcompsoc
\usepackage[caption=false,font=normalsize,labelfont=sf,textfont=sf]{subfig}
\else
\usepackage[caption=false,font=footnotesize]{subfig}
\fi

\def\BibTeX{{\rm B\kern-.05em{\sc i\kern-.025em b}\kern-.08em
    T\kern-.1667em\lower.7ex\hbox{E}\kern-.125emX}}
\usepackage[normalem]{ulem}
\usepackage{multirow}
\usepackage{booktabs} 
\usepackage{xspace}
\usepackage{graphicx}
\usepackage[colorinlistoftodos]{todonotes}
\usepackage{siunitx}
\usepackage{comment}

\newcommand{\cvf}{\ensuremath{cvf}\xspace}

\newcommand{\br}[1]{\ensuremath{\langle #1\rangle}\xspace}
\newcommand{\actnode}{active-node\xspace}
\newcommand{\pasnode}{passive-node\xspace}

\newcommand{\duongtodo}[1]{{\color{red}{{\color{blue}[TODO:} #1]}}}

\newcommand{\sandeeptodo}[1]{{\color{purple}{{\color{blue}[TODO:} #1]}}}

\newcommand{\eves}{\textit{EVE-S}\xspace}
\newcommand{\eveas}{\textit{EVE-AS}\xspace}

\newcommand{\rollback}{\textit{Rollback}\xspace}

\newcommand{\seq}{\textit{SEQ}\xspace}

\newcommand{\maxmatch}{\textit{MAX-MATCH}\xspace}
\newcommand{\coloring}{\textit{COLOR}\xspace}
\newcommand{\pcoloring}{\textit{P-COLOR}\xspace}

\begin{document}

\title{Technical Report: \\
Benefits of Stabilization versus Rollback in Self-Stabilizing Graph-Based Applications on Eventually Consistent Key-Value Stores
\thanks{This work is supported by NSF XPS 1533802.}
}

\author{\IEEEauthorblockN{Duong Nguyen}
\IEEEauthorblockA{\textit{Department of Computer Science and Engineering} \\
\textit{Michigan State University}\\
nguye476@cse.msu.edu}
\and
\IEEEauthorblockN{Sandeep S. Kulkarni}
\IEEEauthorblockA{\textit{Department of Computer Science and Engineering} \\
\textit{Michigan State University}\\
sandeep@cse.msu.edu}
}

\maketitle


\begin{abstract}
In this paper, we evaluate and compare the performance of two approaches, namely self-stabilization and rollback, to handling consistency violating faults (\cvf) that occur when a self-stabilizing distributed graph-based program is executed on an eventually consistent key-value store.
Consistency violating faults are caused by reading wrong values due to
weaker level of consistency provided by the key-value store. 
One way to deal with these faults is to utilize rollback whereas another way is to rely on the property of self-stabilization that is expected to provide recovery from arbitrary states. 
We evaluate both these approaches in different case studies --planar graph coloring, arbitrary graph coloring, and maximal matching-- 
as well as for different problem dimensions such as input data characteristics, workload partition, and network latency. We also consider the effect of executing non-stabilizing algorithm with rollback with a similar stabilizing algorithm that does not utilize rollback.

\end{abstract}


\section{Introduction}
\label{sec:intro}







Consider a computation problem on a large graph (for example matching, coloring, clustering, triangle/truss enumeration, etc. \cite{GK2010JPDC,truss.cohen2008NSA-TR}) where the task is to identify properties of graph structural elements (e.g. nodes, edges) that satisfy some constraints. 
%
For example, in the graph coloring problem, the goal is to find a color assignment for every graph node such that neighboring nodes have different colors and the number of colors used should be as small as possible. These problems have a wide range of immediate applications in practice such as scheduling, resource allocation, 
college student placement~\cite{BS1999JET},
or could be used as sub-procedures in other tasks such as complex network analysis 
\cite{LNR2017Cambridge}.


To improve performance when processing large graphs, the task is partitioned among multiple clients (workers). Coordination among clients could be message passing~\cite{MWM.thinkvertex.2015ACMSurvey} or shared memory~\cite{GraphLab2012VLDB,Lynch96}. In the latter paradigm, the state of the whole graph can be saved in a key-value data store and the clients operate on it~\cite{Trinity2013SIGMOD}. Specifically, a typical client step to process a node includes reading the state of that node (and its neighbors) and updating the node state. For example, in graph coloring problem the client reads the colors of the node and its neighbors, selects a new color distinct from all the neighbors, and updates the node with the new color.
(To reduce communication overhead, the client may process multiple nodes in one step.) For the algorithm to be correct, it is critical that the values read by the clients are up-to-date. This requirement stipulates that: (1) the clients use some mechanisms such as locking to ensure action atomicity~\cite{Dolev2000} and (2) the data store is sequentially consistent~\cite{Lamport79TC}. The first property helps clients avoid reading unreliable values which could lead to incorrect results. 
For example, without action atomicity, two clients may simultaneously update the color values of two neighboring nodes from 0 to 1 and the resultant coloring are invalid.
The second property provides the clients with a consistent view of every data item. Without it, two clients may observe, for example, the status of a shared lock differently and access the critical section simultaneously.

The second property is automatically achieved if the data store contains only one copy (replica) of the data.
However, most data stores in practice~\cite{Dynamo.DHJKLPSVV07SOSP,Cassandra.LM10SIGOPS,Voldemort,DKVF.RK18ASE} consist of multiple replicas for reasons of fault-tolerance, improved access time, and availability of data. Enforcing sequential consistency across multiple replicas, especially in the presence of transient faults (which is a norm in large distributed systems), results in significant delay from the client perspective. This poor performance makes sequential consistency inadequate for many applications~\cite{Dynamo.DHJKLPSVV07SOSP}.
A weaker consistency protocol such as eventual consistency~\cite{Vogels2009CACM} substantially improves the throughput but at the risk of faults where clients update the data erroneously because they read stale values.
This kind of error is denoted as consistency violating fault (\cvf)~\cite{NKD2019ICDCN}. Preventing \cvf{s}, i.e., preventing access to stale data, essentially requires sequential consistency and thereby retards the performance.

One approach to handle \cvf is to have a monitor running concurrently with the program to detect instances of \cvf at runtime.
When a violation (\cvf) is detected, the program is restored to a previous state that is correct and continues its execution thereafter. In~\cite{NCKD2019JBCS}, the authors demonstrated that for some graph-based applications the detect-rollback approach (or rollback for brevity)
works
since \cvf{s} are not frequent and the detection and recovery could be done efficiently.
%
An alternate approach is to use self-stabilization~\cite{EDW426} (or stabilization for brevity). A self-stabilizing (stabilizing) program is guaranteed to recover from an arbitrary state to a legitimate state. Thus, if a client ends up executing a \cvf (updating the information of some node based on stale information), 
the \cvf can be treated as if a fault caused the state of that node to be perturbed. A stabilizing program is designed to recover from such a fault as long as such faults do not occur frequently. (We refer the reader to Sections~\ref{sec:cvf} and \ref{sec:detect-rollback} for more details about stabilization and rollback.)
%
%

\textbf{Summary of the main results.} 
In this paper, we 
investigate
the benefits of the two approaches for handling \cvf{s}.
Clearly, if the underlying program is not stabilizing then we must 
rely on the rollback approach.
Hence, we 
focus on stabilizing programs where both approaches are 
applicable.
Specifically, we 
consider three stabilizing graph-based problems/programs: planar graph coloring, arbitrary graph coloring, and maximal matching.
We run experiments on LinkedIn's Voldemort key-value store on our local network and Amazon Web Service (AWS) network (source code and experimental results are available at \cite{Nguyen20ZenodoICDCS}) and obtain the following observations: 
\begin{itemize}
    \item Stabilization approach provides higher benefits than the rollback approach in the case studies used in this paper. Using sequential consistency as the base-line for comparison, stabilization improves the convergence time of the programs by 25~\% to 35~\%, whereas the rollback approach improves the convergence time by 30~\% in the best case and potentially causes performance to suffer.
    \item 
    We observe a sizeable proportion of the computation time is spent on obtaining locks by the clients to ensure the action atomicity requirement is satisfied (clients do not update the states of neighboring nodes simultaneously-- cf. Section~\ref{sec:distprogram}). In stabilization, we take a further step removing such locks from the clients and treats violations of action atomicity as additional \cvf{s}. This aggressive stabilization approach 
    eliminates the locking overhead at the cost of extra \cvf{s}.
    Experimental results show that the convergence time of the programs speeds up by 2 to 15 times with aggressive stabilization, which suggests that the stabilization cost for the extra \cvf{s} is outweighed by the benefits of no locking overhead.
    \item We analyze the \cvf{s} caused by the 
    absence of locks and find that many of those \cvf{s} resolve favorably by themselves (they do not result in erroneous computation), 
    thus reducing the actual stabilization cost. 
    In contrast, the removal of locks would require rollback approach to utilize more complicated mechanisms to detect atomicity violation instances. The overhead of such a mechanism is expensive, thus preventing an aggressive rollback approach.
    \item Although being more beneficial, aggressive stabilization could suffer from some \cvf{s} that prevent the programs to converge. We propose some heuristics to improve the performance of stabilization in such cases.
    \item We analyze the performance of both approaches under different dimensions such as types of problems, characteristics of input graphs, partitioning schemes, and network latency. We observe that for most of these factors, the impact of a factor on stabilization and rollback is different.
\end{itemize}

\textbf{Contributions of the paper.} To the best of our knowledge, our paper is the first to compare the benefits of stabilization and rollback in handling 
\cvf{s} that occur during the execution of graph-based programs on eventually consistent key-value stores. We find that when the stabilization option is available, it usually provides better benefits than the rollback option. Moreover, stabilization yields its best performance if we aggressively disable mechanisms for action atomicity and treat the violations as additional \cvf{s}. However, for some problems stabilizing algorithms do not exist. In such circumstances, the rollback approach may be the choice. We also analyze different factors that affect the performance of both approaches. This analysis may be informative for designers to obtain a better decision for their problems.

\textbf{Organization of the paper.} In Section \ref{sec:sys-model-architecture}, we present the system models/architecture, the definition of \cvf, and briefly recall the stabilization and rollback approaches. Section~\ref{sec:exp-setup} describes the experiment setup.
In Section~\ref{sec:exp-result}, we compare and analyze the performance of stabilization and rollback. 
Section~\ref{sec:discussion-experiment-results} analyzes the experiment results and their implications. 
Section~\ref{sec:related-work} is related work. We conclude the paper in Section~\ref{sec:concl}.
Due to reason of space, we are only able to briefly discuss some contents in the main paper and provide more detailed discussion in the Appendix.


\section{System Model/Architecture}
\label{sec:sys-model-architecture}

In this section, we recall some important notions used in this paper that have been introduced in \cite{NKD2019ICDCN, NCKD2019JBCS}. 
Section \ref{sec:distprogram} relates the graph algorithms from Introduction to model of distributed programs. Specifically, the \pasnode model in Section \ref{sec:distprogram} is related to graph algorithms in a straightforward manner. We also discuss the \actnode model that is typically used in distributed programs and identify the relation between them.
In Section \ref{sec:voldemort-kv}, we describe the architecture of the Voldemort key-value store and how it implements the \pasnode model. Next, we describe consistency violating faults (\cvf) which are caused by data anomalies in eventual consistency. 
Section \ref{sec:stabilization} and \ref{sec:detect-rollback} respectively recalls the notion of stabilization and detection-rollback for handling \cvf.

\subsection{Distributed Programs: Active and Passive Node Model}
\label{sec:distprogram}
A program $p$ consists of a set of nodes $V_p$ and a set of edges $E_p$. We assume that $\forall i \in V_p, (i,i) \in E_p$. Each node, say $j$, in $V_p$ is associated with a set of variables $var_j$. The union of all node variables is the set of variables of the program $p$, denoted by $var_p$.
A state of $p$ is obtained by assigning each variable in $var_p$ a value from its domain. 
State space of $p$, denoted by $S_p$, is the set of all possible states of $p$. 

Each node $j$ in program $p$ is also associated with a set of actions $ac_j$. An action in $ac_j$ is of the form $g \longrightarrow st$, where the guard $g$ is a predicate involving $\{ var_k :  (j, k) \in E_p \}$ and $st$ updates one or more variables in $var_j$.
We say that an action $ac$ (of the form $g \longrightarrow st$) is enabled in state $s$ if and only if $g$ evaluates to true in state $s$. A node $j$ is said to be enabled at state $st$ if any action in $ac_j$ is enabled in $st$.
The transitions of action $ac$ (of the form $g \longrightarrow st$) are given by  $\{ (s_0, s_1) |$  $s_0, s_1 \in S_p$, $g$ is true in $s_0$ and $s_1$ is obtained by execution $st$ in state $s_0$\}. 
Transitions of node $j$ (respectively, program $p$) is the union of the transitions of its actions (respectively, its nodes). We use $\delta_{ac}, \delta_j$ and $\delta_p$ to denote the transitions corresponding to action $ac$, node $j$ and program $p$ respectively. 
%


\textbf{Computation in traditional/active node model. }
In the traditional/\actnode model, the computation program $p$ is of the form $\br{s_0, s_1, \cdots}$ where 

\begin{itemize}
\item  $\forall l: l \geq 0: $ $s_l$ is a state of $p$,
\item $\forall l: l \geq 0: (s_l, s_{l+1})$ is a transition of $p$ 
or ($(s_l = s_{l+1})$ and no action of $p$ is enabled in state $s_l$), and
\item If some action $ac$ of $p$ (of the form $g \longrightarrow st$) is continuously enabled (i.e., there exists $l$ such that $g$ is true in every state in the sequence after $s_l$) then $ac$ is eventually executed (i.e., for some $x \geq l$, $(s_x, s_{x+1})$ corresponds to execution of $st$.) 

\end{itemize}

The above \actnode model assumes that there is one process residing at each graph node to execute that node's actions, thus this model is likely not suitable for large graphs.
The graph algorithms discussed in Section \ref{sec:intro} correspond to the \pasnode model \cite{NKD2019ICDCN}. Intuitively, the nodes in the graph algorithms correspond to the nodes in the distributed system ($V_p$). The variables associated with nodes are stored in a key-value store.
Specifically, variables of node $k$ are stored as a pair $\br{k,v}$, where $v$ denotes  variables of node $k$.


%

The actions associated with nodes are executed by \textit{clients}. Specifically, in \pasnode model, the system contains a set of clients. Each client is assigned (either statically or dynamically) a subset/partition of the whole set of nodes $V_p$. Each client is responsible for the execution of the actions of enabled nodes assigned to it. 

To execute an action of node $k$, a client (responsible for node $k$) reads the relevant values of variables required to perform the action and updates the relevant variables of node $k$.  
%
%
As an illustration, a graph coloring algorithm would read the colors of $k$ and its neighbors and update color of $k$ if 
the current color of $k$ conflicts with any of the neighbors.

\textbf{Computation in the passive node model. }
The notion of computation in passive-node model is identical to that of active-node model given above; the only difference is that we require clients to execute actions of each node assigned to them in a fair manner, where each node --that has an action enabled-- is executed infinitely often. 

\textbf{(Implicit) Atomicity requirement in \actnode 
and \pasnode model. }
Both the active and the passive node model (implicitly) assume atomicity requirements \cite{Dolev2000,CWHL2013AISA}
which are captured by `$(s_l,s_{l+1})$ is a transition of $p$'. To execute this transition, the program needs to atomically read/write the relevant variables involved in that transition. In \actnode model, one way to achieve this is via local mutual exclusion so that when one node is executing its action its neighbors are not executing their actions. 
In \pasnode model, the atomicity is achieved if 
it is ensured that 
(1) two  clients are not updating state of neighboring nodes simultaneously, and (2) each client receives the most updated values of nodes it operates on.
The first requirement can be realized by locks to avoid simultaneous client updates and we denote the time the clients spend to ensure this requirement as \textit{locking coordination overhead} (or locking overhead). (We refer to Appendix Section~\ref{sec:graph-locks} for details of lock implementation in this paper.) The second requirement is fulfilled when the key-value store is sequentially consistent.


\color{black}

\color{black}

\subsection{Voldemort Key-Value Store}
\label{sec:voldemort-kv}

As discussed in Section \ref{sec:distprogram}, in the \pasnode model the variables of all nodes are stored in a key-value store. In this paper, we use Voldemort --an open-source implementation of Amazon Dynamo \cite{Dynamo.DHJKLPSVV07SOSP}-- to implement the \pasnode model.

{While more details of Voldemort operation is provided in Appendix Section~\ref{sec:voldemort-operation}, essentially Voldemort is an active-replication-based key-value store. The key parameters of active replication are the number of replicas ($N$), the number of required replies for read request ($R$), and the number of required confirmations for write request ($W$). By adjusting the values of $W, R$, and $N$, the consistency model of the key-value store is changed. For example, if $W+R > N$ and $W > \frac{N}{2}$ then the consistency is sequential. If $W + R \le N$ then it is eventual consistency.}


\subsection{Consistency Violating Faults (\cvf)}
\label{sec:cvf}



In sequential consistency, clients always obtain the fresh data whereas in eventual consistency, clients may read a stale value due to transient faults.
As a result of such violation in eventual consistency, the computation of the given program $p$ is a sequence $\br{s_0, s_1, \cdots}$ such that  most transitions $(s_l, s_{l+1}), l \geq 0$ in this sequence belong to $\delta_p$ (the set of transitions of $p$) and some transitions correspond to the scenario where some client working on node $j$ reads a stale value of some variable and (incorrectly) updates one or more variables of $j$. This faulty transition is effectively the same as perturbing one or more variables of node $j$. We denote such transitions as consistency violating faults ($\cvf_p$) and, by the above discussion, $\cvf_p$ is a \textit{subset} of 
%
$\{ (s_0, s_1) | s_0, s_1 \in S_p$ and $s_0$, $s_1$ differ only in the variables of some node $j$ of $p\}.$

\textbf{Remark 1. }
Whenever program $p$ is clear from the context, we use $\cvf$ instead of $\cvf_p$. 

\textbf{Remark 2. } If the clients do not utilize a mechanism to guarantee atomicity, they may read 
unreliable 
values that are not supposed to be read (values being updated by other clients), and incorrectly calculate values for some variables of some node, e.g. node $j$. When these incorrect values are updated to the store, that update has the same effect as perturbing the variables of node $j$. In other words, the incorrect transitions caused by violations of atomicity requirements can also be treated as \cvf{s}.

For brevity reason, we omit some discussion and examples of \cvf{s} and refer reader to Appendix Section~\ref{sec:cvf-more} for those contents.

\subsection{Stabilization}
\label{sec:stabilization}

In this section, we recall the definition of self-stabilization (or stabilization) from \cite{EDW426}. Using the definition of computation from the previous section, stabilization is defined  as follows:


\textbf{Stabilization. }
Let $p$ be a program. Let $I$ be a subset of state space of $p$. We say that $p$ is stabilizing with state predicate $I$ ($I$ is denoted as the invariant of $p$) iff 


\begin{itemize}
\item \textit{Closure:} If program $p$ executes a transition in a state in $I$ then the resulting state is in $I$, i.e., for any transition $(s_0, s_1) \in \delta_p$, $s_0 \in I \Rightarrow s_1 \in I$, and 
\item \textit{Convergence:} Any computation of $p$ eventually reaches a state in $I$, i.e., for any $\br{s_0, s_1, \cdots}$ that is a computation of $p$, there exists $l$ such that $s_l \in I$. 
\end{itemize}

A special case of stabilization is \textit{silent stabilization} where once the program reaches $I$, 
it has no enabled actions, and thus, the program will remain in that state forever (unless perturbed by faults). This paper focuses only on such \textit{silent stabilizing} programs. We refer the reader to \cite{NKD2019ICDCN} for discussion of non-silent stabilizing programs.  

%

\textbf{Silent Stabilization. }
Let $p$ be a program. Let $I$ be a subset of state space of $p$. We say that $p$ is silent stabilizing with state predicate $I$ iff 

\begin{itemize}
\item \textit{Closure:} Program $p$ has no transitions that can execute in $I$, i.e., for any $s_0 \in I$, $(s_0, s_1) \not \in \delta_p$ for any state $s_1$, and 
\item \textit{Convergence:} Any computation of $p$ eventually reaches a state in $I$, i.e., for any $\br{s_0, s_1, \cdots}$ that is a computation of $p$, there exists $l$ such that $s_l \in I$. 
\end{itemize}



\color{black}

\textbf{Stabilization of programs in the presence of \cvf. } As discussed in Section~\ref{sec:cvf}, the effect of a \cvf is the same as perturbing one or more variables of a node. Moreover, \cvf{s} are by design not deliberate and not frequent~\cite{Dynamo.DHJKLPSVV07SOSP}. Thus, a stabilizing program is likely to have the opportunity to execute several valid transitions between two \cvf{s}. Although some specific \cvf perturbations may significantly prolong the convergence of the program, the likelihood of such perturbations is small. Therefore, it is expected that a program in eventually consistent \pasnode model still stabilizes despite the additional overhead of correcting \cvf{s} in exchange for the higher performance of a weaker consistency.

\subsection{Detect Rollback Approach}
\label{sec:detect-rollback}

In this section, we briefly recall the detect-rollback approach (or rollback for brevity) to handle \cvf. 
In the rollback approach, the user provides a correctness property $\Phi$ that the computation should always satisfy. We note that $\Phi$ can be the conjunction of smaller correctness properties, i.e. $\Phi = \bigcap_{i}\Phi_i$.
The user runs the program on eventually consistent \pasnode model as well as the monitors. During the execution of program $p$, if property $\Phi$ is violated (any $\Phi_i = $ \si{false}) due to the occurrence of \cvf, the monitors will detect such violations and inform the computation to roll back to the most recent state where $\Phi$ is satisfied, and the computation is resumed from there.
In the problem associated with violation of \cvf{s}, the monitory for detecting predicate $\Phi_i$ is semi-linear and we use the algorithm in \cite{CG98DC}. 
{We note that a \cvf involves the reading of stale information of a node by at most two clients. This locality property allows the program to undo the effect of the \cvf by just restarting a few actions by the clients involved in the \cvf rather than requiring all clients to rollback in a coordinated fashion.}
A more detailed description of the monitor algorithm and rollback approach is provided in~\cite{NCKD2019JBCS}.

%

\color{black}



\section{Experiment Setup}
\label{sec:exp-setup}

\subsection{System Configuration}
\label{sec:sys-config}

We ran experiments in two environments: local lab network and Amazon Web Service (AWS) network. 
In the local lab network, we are able to control the network latency between the clients and servers by using proxies. On the other hand, the AWS network provides an environment similar to a realistic network and the network latency is determined by actual network conditions. More details of the machine and proxy configuration are provided in the Appendix, Section~\ref{sec:sys-config-more}.

In our experiments, the distributed system consisted of 3 regions (clusters). Each region had 1 server machine (which hosted 1 replica) and 2 client machines (a client machine hosted 5 client processes). Thus, there were 3 servers and 30 clients. We chose the configuration N3R1W1 (number of replicas N=3, number of required reads R=1, number of required writes W=1) for eventual consistency, and N3R1W3 for sequential consistency (in our experiments, N3R1W3 performed better than another sequential consistency configuration N3R2W2).

\subsection{Client Execution Modes.} 
\label{sec:client-execution-modes}

\begingroup 
\renewcommand{\arraystretch}{1.1} 
\begin{table}[t]
\caption{Four client execution modes}
\vspace{-5pt}
\begin{center}
\begin{tabular}{|p{1.0cm}|p{1.0cm}|p{1.4cm}|p{1cm}|p{2.2cm}|}
    \hline
    Execution mode  & Consis-tency   & Atomicity mechanisms & Monitors & Note \\
    \hline
    \seq    & Sequential    & Yes       & No        & No \cvf. Standard approach. \\
    \eves   & Eventual      & Yes       & No        & Infrequent \cvf expected. \\
    \eveas  & Eventual      & No        & No        & More \cvf expected. \\
    \rollback & Eventual     & Yes       & Yes       & Rollback when violation is detected. \\
    \hline
\end{tabular}
\label{table:client-execution-mode}
\end{center}
\end{table}
\endgroup

The clients were configured to run in four different modes (cf. Table \ref{table:client-execution-mode}) corresponding to four different ways of executing the computation. 
In \textit{sequential} mode (\seq), the clients run on sequentially consistent key-value store and use mechanisms (e.g. locks) to guarantee atomicity. 
No \cvf{s} occur in SEQ mode. This is the standard approach for executing the computation~\cite{MWM.thinkvertex.2015ACMSurvey,GraphLab2012VLDB} and is used as the baseline for comparison.
In \textit{eventual with stabilization} mode (\eves), the clients also employ 
mechanisms for atomicity
but run on eventually consistent data store. This mode allows \cvf to occur due to eventual consistency. However, \cvf is expected to be infrequent so that between two instances of \cvf, the clients can execute several transitions to stabilize the computation. 
\textit{Eventual with aggressive stabilization} mode (\eveas) is similar to 
\eves except that the clients do not use 
mechanisms for atomicity. Consequently, in \eveas more \cvf{s} are expected (Remark 2) but the locking overhead is avoided. 
Lastly, in \rollback mode, the clients run on eventually consistent data store and also use 
atomicity mechanisms.
Hence, \cvf occurs in rollback mode. However, instead of relying on the stabilizing transitions of the program to correct \cvf, the monitors are deployed to detect violations and the computation is then rolled back to undo the effect of \cvf.


To compare the performance of different execution modes, we use convergence time 
as the measurement. We note that for a silently stabilizing program, 
convergence time is the time it terminates.
We refer to Appendix Section~\ref{sec:termination-detection-algo} for the description of termination detection algorithm.

\subsection{Case Study Problems}
\label{sec:exp-case-studies}
We used three stabilization problems/programs as our case studies: arbitrary/general graph coloring (\coloring), planar graph coloring (\pcoloring), and maximal matching (\maxmatch).
%
For \coloring, we used the stabilizing algorithm in \cite{GT2000OPODIS} (the first of three variations) and noted that a perturbed state of a node (caused by \cvf) in \coloring can be corrected by just one action.
%
%
For \pcoloring, we implemented the algorithm in~\cite{GK93DC} that uses at most 6 colors. Unlike \coloring algorithm, \pcoloring algorithm consists of two steps (which can run simultaneously):  constructing a directed acyclic graph (DAG) and coloring the nodes based on that DAG.
%
%
We used the algorithm in ~\cite{MMPT2009TCS} to find the maximal matching of a graph. In \maxmatch, a \cvf may require several actions to correct.

\subsection{Input Graphs}
\label{sec:input-graphs}

We used three types of input graphs in the experiments: planar graphs, social graphs, and random regular graphs. A planar graph is a graph that can be drawn on a plane such that its edges do not cross with each other. 
We used the algorithm and program in \cite{Fusy2009RSA} to generate planar graphs of approximately 10,000 nodes (and roughly 24,000 edges).
In a social graph, node degrees follow the power-law distribution and nodes form clusters within the graph. In a random regular graph, nodes have the same degree and are randomly connected.
We used the tool \textit{networkx} \cite{networkx} to generate social and random regular graphs. These graphs had 10,000 to 50,000 nodes.

\subsection{Workload Partitioning Schemes.}
\label{sec:partitioning-scheme}

In the \pasnode model, each client is responsible for a (roughly equal) partition of the graph. We used three schemes to construct the clients' partitions. In the normal partitioning (or straight partitioning), each client is responsible for a trunk of consecutive nodes.
For example, with 10,000 nodes and 10 clients, client 0 is assigned nodes 0 to 999, ..., client 9 is assigned nodes 9,000 to 9,999.
In the Metis partitioning, we used graph partitioning tool Metis \cite{A-RK2006IPDPS} to partition the graphs. Metis partitioning algorithm aims to minimize the \textit{edge-cut} partitioning objective, i.e. the number of graph edges bridging different partitions, and thus increases the locality within the partitions. 
In the random partitioning, each client is assigned a distinct set with roughly the same number of nodes randomly selected from the graph. 
Random partitioning distributes the workload more evenly between clients but could have negative effect on the locality of partitions.

\section{Benefits of Stabilization versus Rollback: Comparison and Analysis}
\label{sec:exp-result}


\subsection{Stabilization vs. Rollback: Comparison and Analysis}
\label{sec:result-basic:ss-vs-rollback}

\textbf{Overall comparison. }
Table \ref{tab:self-stabilization-vs-rollback} shows the experiment results of running four execution modes (cf. Section \ref{sec:client-execution-modes}) on different case study problems and input graphs. Input graphs are partitioned with normal partitioning scheme. We ran experiments in local lab network where the average latency was 20 \si{\ms} (using proxy) and 
each measurement is the average of several runs.
The sequential mode (\seq) is used as the baseline of comparison.



In general, stabilization performed better than rollback in our case studies. Specifically, stabilization \eves improved the convergence time by 25\%--35\% whereas \rollback improved the convergence time by 29\% in the best case but potentially caused the performance to suffer. Remarkably, aggressive stabilization \eveas improved the performance 2--15 times.

\begin{table*}[htbp]
\centering
\caption{Stabilization vs. Rollback.
Graphs are partitioned in normal scheme. Network latency was 20 \textit{ms}.
\seq is baseline for comparison. 
Rows 7-10 are convergence time benefits, shown in percentage increase or in speedup (e.g. $\times 5.2$ means 5.2 times faster). 
}
\begin{tabular}{|p{1.5cm}|p{2.8cm}|r|r|r|r|r|r|} 
\hline
& {\parbox{1.5cm}{Problem}} & \multicolumn{1}{c|}{\parbox{2.2cm}{Planar Graph Coloring (\pcoloring)}} & \multicolumn{2}{c|}{\parbox{2.4cm}{Arbitrary Graph \\Coloring (\coloring)}} & \multicolumn{3}{c|}{\parbox{3.6cm}{Maximal Matching \\(\maxmatch)}} \\ \hline
& Input graph & Planar 10K & Social 50K & Regular 50K & Social 10K & Regular 10K & Planar 10K\\ 
\hline
\multirow{4}{*}{\parbox{1.5cm}{\centering{Convergence time (seconds)}}} & SEQ                   & 3,887 & 27,995 & 6,518 & 31,581 & 14,859 & 8,545 \\ 
& EVE-S                   & 2,658 & 18,229 & 4,270 & 23,246 & 11,028 & 6,173\\ 
& EVE-AS                 & 754  &  1,885 & 3,547 & 2,892 & 1,866 & 2,590 \\ 
& Rollback          & 3,860 & 32,165 & 4,624 & 32,238 & 12,496 & 8,660 \\  
\hline
\multirow{4}{*}{\parbox{1.5cm}{\centering{Benefit}}}& EVE-S vs. SEQ       & 31.6\% & 34.9\% & 34.5\% & 26.4\% & 25.8\% & 27.8\%  \\ 
& EVE-AS vs. EVE-S     & $\times 3.5$ & $\times 9.7$ & $\times 1.2$ & $\times 8$ & $\times 5.9$ & $\times 2.4$\\ 
& EVE-AS vs. SEQ     & $\times 5.2$ & $\times 14.9$ & $\times 1.8$ & $\times 10.9$ & $\times 8$ & $\times 3.3$ \\ 
& Rollback vs. SEQ & 0.7\% & -14.9\% & 29\% & -2.1\% & 15.9\% & -1.4\% \\ 
\hline
\end{tabular}
\label{tab:self-stabilization-vs-rollback}
\end{table*}


\textbf{Impact of input graph structure.} The structure of input graph affects the computation in two ways: (1) it changes the work balance between clients and (2) it determines the locking overhead among the clients. 

In skewed graphs such as social graphs and planar graphs where there are a few nodes with very high connectivity degrees, some clients will be assigned graph partitions with more work (the number of nodes is roughly the same but the number of edges in these partitions is higher). In contrast, the workload can be evenly distributed in random regular graph due to its regularity structure.

\begin{figure}[!t]
    \centering
    \subfloat[Social graph]{
        \includegraphics[width=0.45\textwidth]{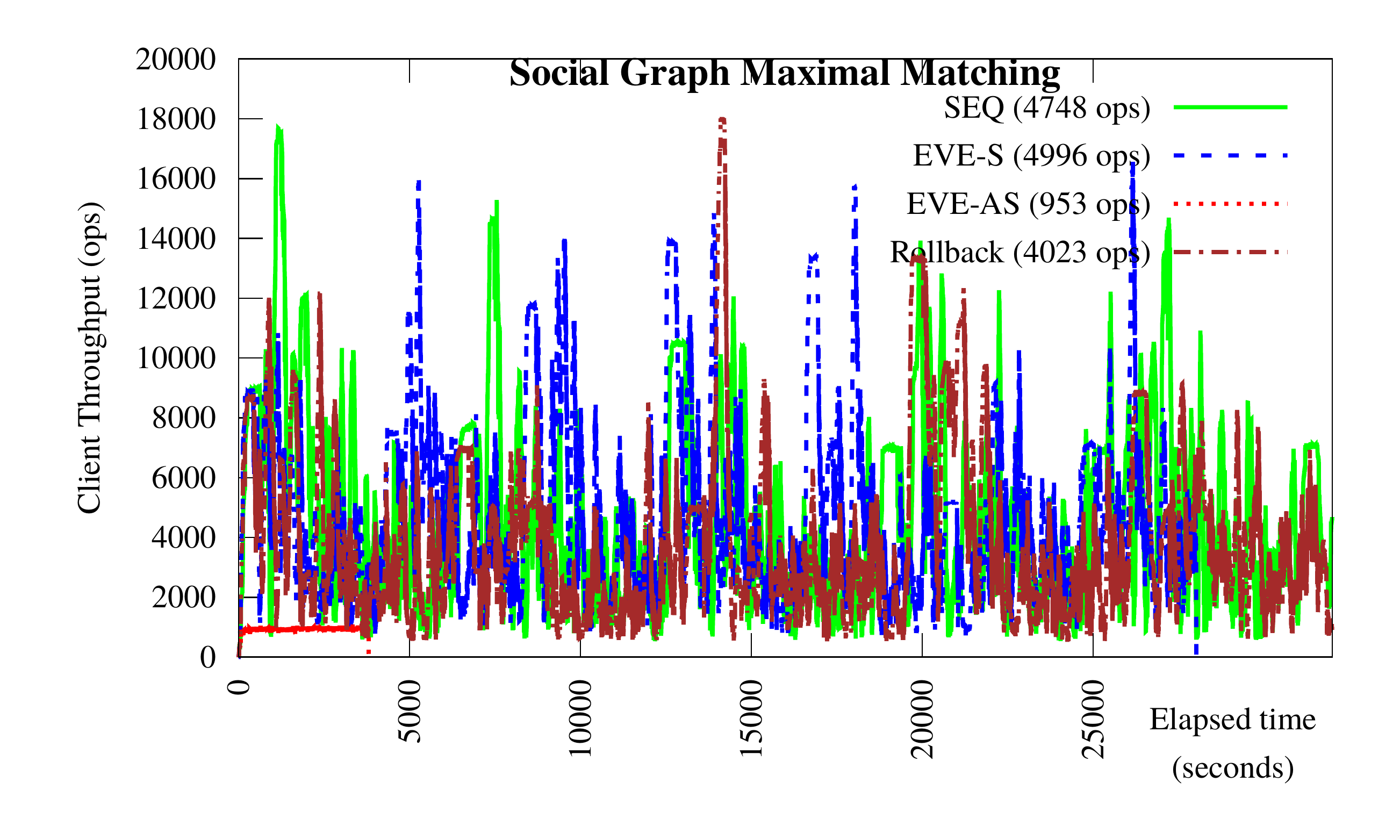}
        \label{fig:max-matching-throughput-powerlaw}}
    \\
    \vspace{-12pt}
    \subfloat[Random Regular graph]{
        \includegraphics[width=0.45\textwidth]{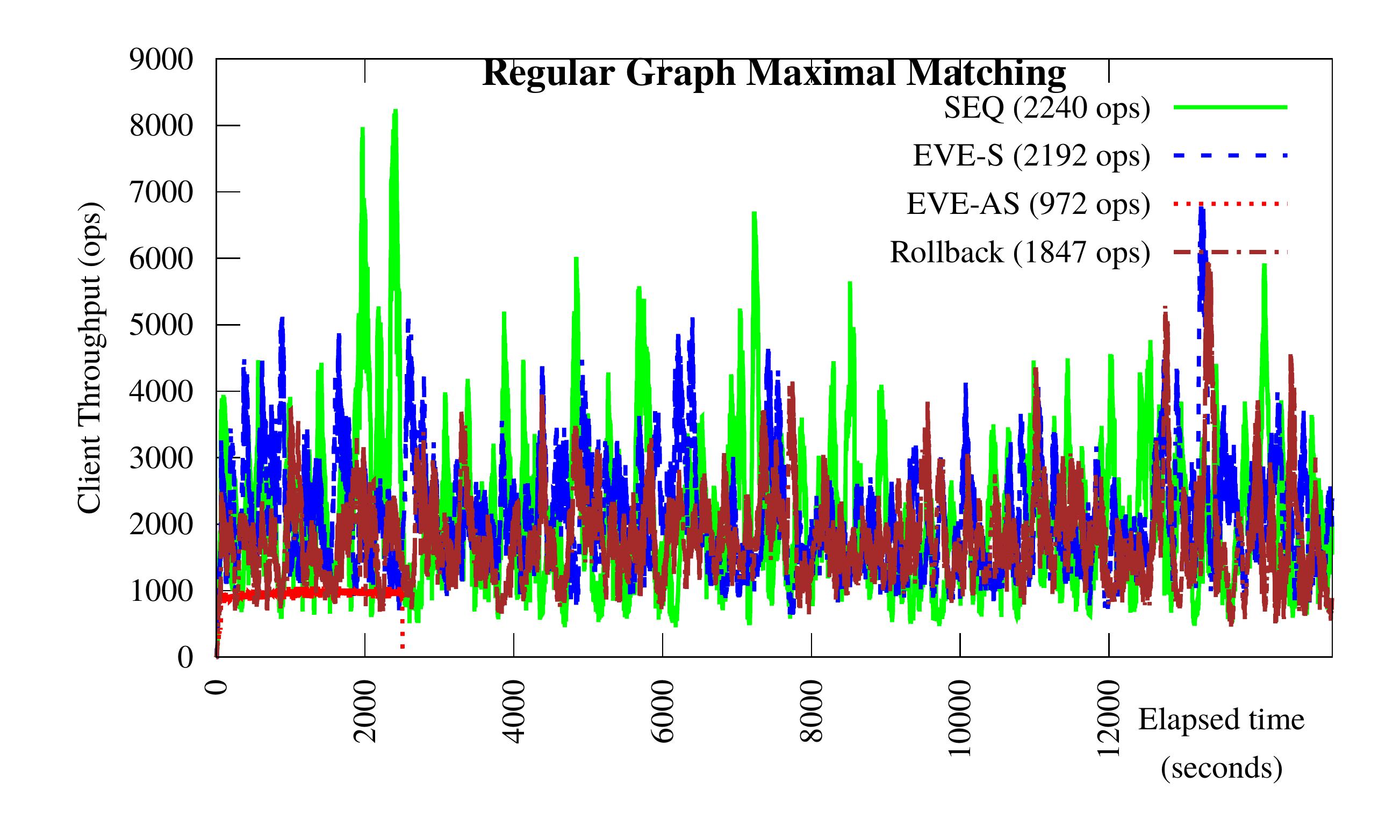}
        \label{fig:max-matching-throughput-regular}}
    \\ 
    \vspace{-12pt}
    \subfloat[Planar Graph]{
        \includegraphics[width=0.45\textwidth]{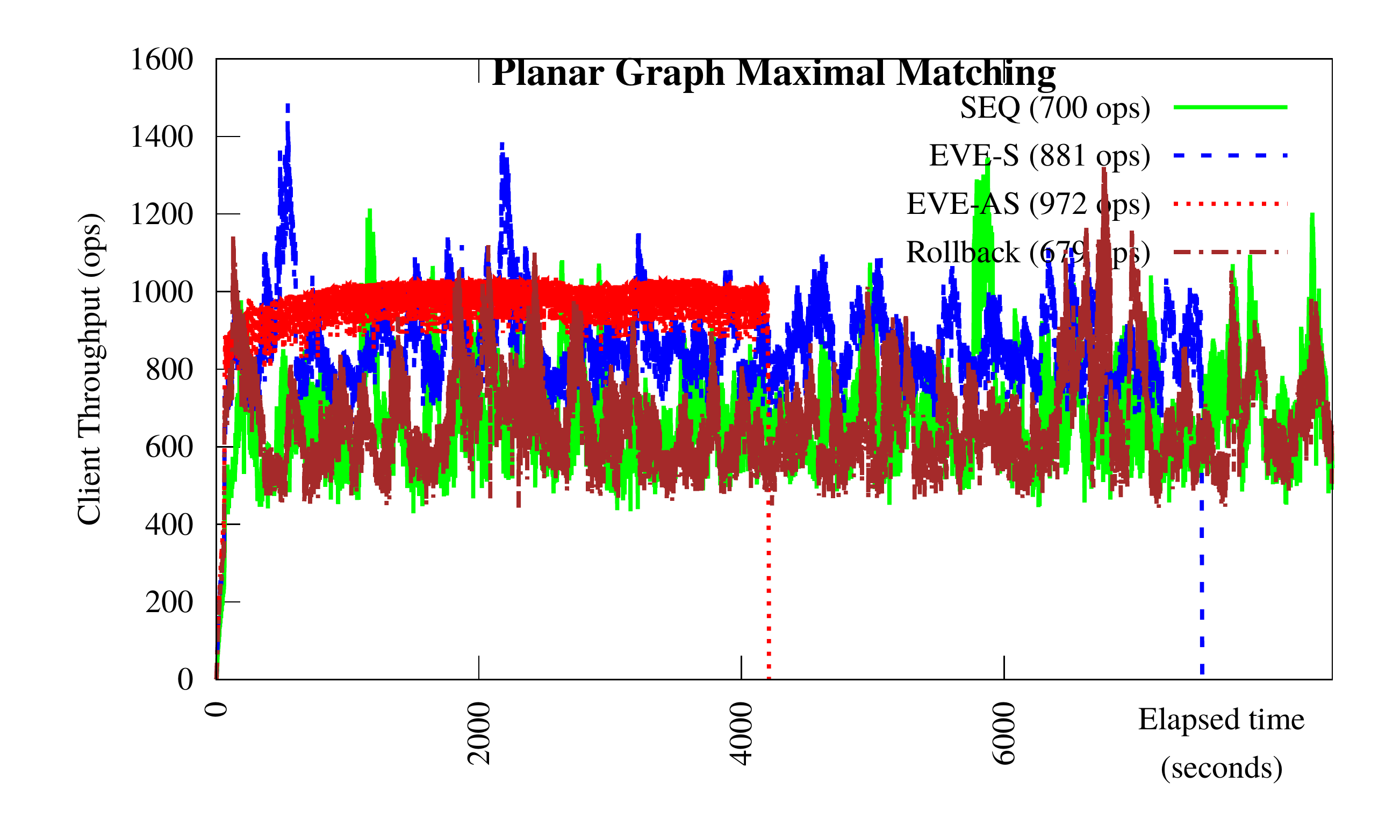}
        \label{fig:max-matching-throughput-planar}}
    \vspace{-3pt}
    \caption{Measurement of client throughput (\textit{ops} -- operations per seconds) of \maxmatch with different input graphs. Normal partitioning. Latency was 20 \textit{ms}.}
    \label{fig:max-matching-throughput}
\end{figure}

The locking overhead 
also depends on the connectivity structure of input graphs. 
%
For example, Figures \ref{fig:max-matching-throughput} (a-c) measures the average throughput of \maxmatch running on 4 execution modes and in different graphs. 
%
%
In social graphs (cf. Figure \ref{fig:max-matching-throughput-powerlaw}), the throughput in \eves (4996 \textit{ops}) was about 5 times higher than that in \eveas (953 \textit{ops}).
In regular graphs (cf. Figure \ref{fig:max-matching-throughput-regular}), this difference was about 2 times (2192 and 972 \textit{ops}).
Lastly, the two throughputs were comparable in planar graph (881 and 972 \textit{ops}, cf. Figure \ref{fig:max-matching-throughput-planar}). 
Since the key difference between \eves and \eveas is whether an atomicity mechanism is used or not (with or without locking overhead),
these results indicated that the locking overhead was highest in social graphs due to their complex structure (power-law degree distribution, clustering) and lowest in planar graphs due to their locality property. (We can partition a planar graph into non-overlapping partitions with a small number of border nodes-- nodes connected to other partitions.)

Due to the amount of locking overhead,
computation was slowest on social graphs and (often) fastest on planar graphs. We note that \maxmatch (on \eveas mode) converged faster on random regular graphs (1,866 \si{\second}) than on planar graphs (2,590 \si{\second}) because some clients converged slower than others in planar graphs (due to skewed partitioning results), which increased the overall time of the program.


The graph structure also affects the benefits of stabilization and rollback. Specifically, the benefits of aggressive stabilization \eveas were highest (lowest, respectively) in social graphs (planar graphs, respectively) since the locking overhead was high (low, respectively) and \eveas avoided such overhead. 
In contrast, the performance of \rollback suffered on social graphs because the chance of conflicts (two clients updated neighboring nodes simultaneously) was high and rollback was more frequent. \rollback performed well on random regular graphs since the chance of conflicts was low. We note that for \maxmatch on planar graphs, \rollback was slightly slower than \seq because of the skewed workload. When planar graphs were partitioned using the random scheme, \rollback was 22\% faster than \seq (we discuss the impact of partitioning schemes later in this section). Finally, the benefits of \eves were 
fairly stable across different settings 
(25\%--35\%) because these benefits stemmed from the performance difference between eventual and sequential consistency. 

\color{black}

\textbf{Impact of case study problems.}
The effect of \cvf is not the same for different problems. In \coloring, a \cvf can cause a client to update a node with color similar to one of its neighbors but this error can be fixed by one valid transition (update the node with a different color). Nodes that are more than one hop away are not affected by the \cvf. 
In contrast, in \maxmatch a \cvf can have cascading effect that requires updates at distant nodes. 
As an illustration, suppose four nodes $v_1$, $v_2$, $v_3$, and $v_4$ are on a straight line in that order. Nodes $v_1$ and $v_3$ are matched with $v_2$ (due to \cvf). To correct this error, we can un-match $v_2$ from $v_3$. Since $v_3$ is now free, it can be matched with $v_4$, thus updating the states of both $v_3$ and $v_4$. Therefore, the cost to correct \cvf{s} in \maxmatch is higher and the benefits of \eveas are smaller in \maxmatch than in \coloring ($\times 10.9$ and $\times 14.9$ speedup on social graphs) since \eveas introduces more \cvf{s}. 


We notice an exception: on regular graphs, the benefit of \eveas in \coloring is unusually low ($\times 1.8$ speedup whereas the benefits of \maxmatch  is $\times 8$ speedup). We examined the execution of \coloring and found that eliminating atomicity mechanisms (in \eveas) introduced some \cvf{s} that were difficult to recover. 
{This happened when only a small number of nodes had inconsistent colors and \cvf{s} caused clients to re-visit those nodes again and again. We only observed these \cvf{s} in regular graphs as the workload was split very evenly across clients, thereby leading to a livelock.}
One way to address this problem is using 
random coloring. For more detailed descriptions of these \cvf{s} and related experiment results, we refer to Appendix Section~\ref{sec:result-extra-var}.

\color{black}

\textbf{Impact of partitioning scheme.}
A normal partitioning of skewed graphs (social or planar) causes workload imbalance among clients and high connectivity among partitions (low locality), which increases the computation time as well as affects the benefits of stabilization and rollback.
Efficient partitioning schemes can address these issues. We consider two alternatives: random partitioning and Metis partitioning. The former helps distribute the workload more evenly whereas the later improves the locality.

As shown in Table \ref{tab:self-stabilization-vs-rollback-planar-random-partition}, \rollback was not better than \seq when normal partitioning scheme was used in \maxmatch and \pcoloring because of the uneven workload. However, when random partitioning was employed, the benefits of both rollback and stabilization were significantly improved (cf. Table~\ref{tab:self-stabilization-vs-rollback-planar-random-partition}).
However, the convergence time often increases with random partitioning because this partitioning 
disturbs the locality of planar graphs. For a detailed 
examination of the effects of
normal and random partitioning schemes, we refer to Appendix Section~\ref{sec:compare-normal-random-partition}. 

Metis partitioning \cite{A-RK2006IPDPS} reduces the number of external edges bridging between partitions, and thus improves the locality within partitions. Consequently, the locking overhead was reduced and the convergence time of all execution modes was improved when compared to normal partitioning (cf. Table \ref{tab:self-stabilization-vs-rollback-metis}). Since the locking overhead was reduced, the benefits of aggressive stabilization \eveas decreased.

\color{black}


\textbf{Impact of network latency.} 
Network latency characterizes the geographical distribution of replicas. As shown in Table~\ref{tab:impact-network-latency}, when network latency increased (from 20 \si{\ms} to 50 \si{\ms}), the benefits of stabilization (\eveas) slightly increased. We attribute this result to the fact that the benefits of eventual consistency compared to sequential consistency increase when network latency increases~\cite{NCKD2019JBCS}. On the other hand, the effect of network latency on the benefits of rollback was mixed. We anticipate that the different interaction patterns of rollback with the underlying stabilizing programs is the reason for this variation.

\color{black}

\begin{table}[htbp]
\centering
\vspace{-5pt}
\caption{Effect of Random Partitioning on Stabilization and Rollback.
Rows 2-5 are convergence time. Rows 6-8 are benefits, in percentage increase or in speedup (e.g. $\times 3$ means 3 times faster). Network latency was 20 \textit{ms}}
\begin{tabular}{|p{1.1cm}|p{1.2cm}|r|r|r|r|} 
\hline
& \multirow{2}{*}{\parbox{1.2cm}{Execution mode}} & \multicolumn{2}{c|}{\maxmatch} & \multicolumn{2}{c|}{\pcoloring} \\
\cline{3-6}
&  &  \multicolumn{1}{c|}{\parbox{1cm}{Normal partition}} & \multicolumn{1}{c|}{\parbox{1cm}{Random partition}} & \multicolumn{1}{c|}{\parbox{1cm}{Normal partition}} & \multicolumn{1}{c|}{\parbox{1cm}{Random partition}} \\
\hline
\multirow{4}{*}{\parbox{1.1cm}{\centering{Conver-gence time (seconds)}}} & SEQ & 8,545 & 10,736 & 3,887 & 8,686\\ 
& EVE-S                   & 6,173 & 7,026 & 2,658 & 5,315 \\ 
& EVE-AS                  & 2,590 & 1,448 & 754 & 655 \\ 
& Rollback                & 8,660 & 8,341 & 3,860 & 7,242 \\  
\hline
\multirow{4}{*}{\parbox{1.1cm}{\centering{Benefit}}} & EVE-S vs. SEQ    & 27.8\% & 34.6\% & 31.6\% & 38.8\% \\ 
& EVE-AS vs. SEQ       & $\times 3.3$ & $\times 9.4$ & $\times 5.2$ & $\times 13.3$ \\ 
& Rollback vs. SEQ     & -1.4\% & 22.3\% & 0.7\% & 16.6\% \\ 
\hline
\end{tabular}
\label{tab:self-stabilization-vs-rollback-planar-random-partition}
\end{table}

\begin{table}[htbp]
\caption{Impact of Metis partitioning scheme. Latency was 20 \textit{ms}.
}
\begin{tabular}{|p{2cm}|p{2cm}|r|r|} 
\hline
\multicolumn{2}{|l|}{\parbox{1.5cm}{Problem}} & \multicolumn{2}{c|}{\parbox{3cm}{\maxmatch}} \\ \hline
\multicolumn{2}{|l|}{Input graph} & Planar & Planar \\
\hline 
\multicolumn{2}{|l|}{Partition scheme} & Normal & Metis \\ 
\hline
\multirow{4}{*}{\parbox{1.5cm}{\centering{Convergence time (seconds)}}} & SEQ                   & 8,545 & 2,585 \\ 
& EVE-S                  & 6,173 & 2,389 \\ 
& EVE-AS                 & 2,590 & 2,154 \\ 
& Rollback               & 8,660 & 2,635 \\  
\hline
\multirow{4}{*}{\parbox{1.5cm}{\centering{Benefit}}}& EVE-S vs. SEQ       & 27.8\% & 7.6\% \\ 
& EVE-AS vs. SEQ     & $\times 3.3$ & $\times 1.2$ \\ 
& Rollback vs. SEQ   & -1.4\% & -1.9\% \\ 
\hline
\end{tabular}
\label{tab:self-stabilization-vs-rollback-metis}
\end{table}

\begin{table}[htbp]
\centering
\caption{Impact of network latency. Rows 4-6 are convergence time (in seconds). Rows 7-8 are the benefits, shown in percentage increase or in speedup (e.g. $\times 4.3$ means 4.3 times faster).}
\vspace{-5pt}
\begin{tabular}{|p{1.1cm}|r|r|r|r|r|r|}
\hline
Program & \multicolumn{2}{c|}{\parbox{1.8cm}{\maxmatch}} & \multicolumn{2}{c|}{\parbox{1.8cm}{\coloring}} & \multicolumn{2}{c|}{\parbox{1.2cm}{\pcoloring}} \\
\hline
Input graph & \multicolumn{2}{c|}{\parbox{1.9cm}{Regular 10K, normal partition}} & \multicolumn{2}{c|}{\parbox{1.9cm}{Regular 10K, normal partition}} & \multicolumn{2}{c|}{\parbox{1.9cm}{Planar 10K, normal partition}} \\
\hline
Latency & 20 \textit{ms} & 50 \textit{ms} & 20 \textit{ms} & 50 \textit{ms} & 20 \textit{ms} & 50 \textit{ms} \\
\hline
SEQ & 14,859 & 35,653 & 6,518    & 15,535 & 3,887 & 9,415 \\
EVE-AS & 1,866 & 3,985 & 1,615 & 3,607 & 754 & 1,814 \\
Rollback & 12,496 & 38,657 & 4,742 & 14,113 & 3,860 & 9,057 \\
\hline
EVE-AS vs. SEQ       & $\times 8.0$ & $\times 8.9$ & $\times 4.0$ & $\times 4.3$ & $\times 5.16$ & $\times 5.19$ \\ 
Rollback vs. SEQ     & 15.9\% & -8.4\% & 28.3\% & 9.2\% & 0.7\% & 3.8\% \\ 
\hline
\end{tabular}
\label{tab:impact-network-latency}
\end{table}



\subsection{Experiments on Amazon AWS}
\label{subsec:result-aws}
To confirm the results in a more realistic deployment, we ran experiments on Amazon Web Services (AWS) network. As shown in Table \ref{tab:aws-self-stabilization-vs-rollback}, the AWS results agree with the experimental results on the local lab network (cf. Table~\ref{tab:self-stabilization-vs-rollback}).

\begingroup
\renewcommand{\arraystretch}{1.1} 
\begin{table}[htbp]
\caption{Experiment results on Amazon AWS network.}
\vspace{-5pt}
\begin{tabular}{|p{1.1cm}|p{1.2cm}|r|r|r|}
\hline
\multicolumn{2}{|c|}{\parbox{1.2cm}{Problem}} & \multicolumn{1}{c|}{\parbox{1.2cm}{\pcoloring}} & \multicolumn{1}{c|}{\parbox{1.1cm}{\coloring}} & \multicolumn{1}{c|}{\parbox{1.7cm}{\maxmatch}} \\ \hline
\multicolumn{2}{|c|}{Input graph} & Planar 10K & Social 10K & Regular 10K \\ \hline
\multicolumn{2}{|c|}{Partition scheme} & Random & Normal & Normal \\ 
\hline
\multirow{4}{*}{\parbox{1.1cm}{\centering{Conver-gence time (seconds)}}} 
    & SEQ                   & 10,211 & 21,265 & 6,816 \\ 
    & EVE-S                  & 6,586 & 13,630 & 4,038 \\ 
    & EVE-AS                 & 797 & 2,430 & 413 \\ 
    & Rollback          & 9,575 & 21,718 & 7,625 \\  
\hline
\multirow{3}{*}{\parbox{1.1cm}{\centering{Benefit}}}
    & EVE-S vs. SEQ       & 35.5\% & 35.9\% & 41.7\% \\ 
    & EVE-AS vs. SEQ     & $\times 12.8$ & $\times 8.8$ & $\times 16.5$ \\ 
    & Rollback vs. SEQ   & 6.2\% & -2.1\% & -11.9\% \\ 
\hline
\end{tabular}
\label{tab:aws-self-stabilization-vs-rollback}
\end{table}
\endgroup

\subsection{Overall/Key Observation}

Although the performance of each specific execution mode depends on several factors, in general, \eveas is noticeably efficient, \eves consistently yields substantial benefits whereas \rollback could provide comparable benefits as \eves but also potentially causes performance to suffer.

The above observation poses some questions: (1) what is the reason that makes stabilization, especially aggressive stabilization, more efficient than rollback?
And (2) when one already has a non-stabilization algorithm for a problem at hand, and there exists another stabilizing algorithm which is (relatively) less efficient (on sequential consistency), is it worth considering the stabilizing option? {This question can also be extended for handling the case where adding stabilization to a non-stabilizing program leads to an increase in overhead/computation time. }
We discuss these questions in Section~\ref{sec:discussion-experiment-results}.

\color{black}

\section{Analysis of Results and Their Implications in the Design}
\label{sec:discussion-experiment-results}



\subsection{Insight into Comparison of Stabilization vs Rollback}
\label{sec:evaluate-cvf}


We observe from Table \ref{tab:self-stabilization-vs-rollback} that the performance of self-stabilization is generally better than rollback, particularly in 
\coloring with social graphs.
We anticipate the reason is that the effect of \cvf{s} is resolved differently in the two approaches. As an illustration, consider 
the \coloring program for social graphs.
A \cvf corresponds to the case when the possession time intervals of two clients for a lock overlapped. (This scenario occurs in eventual consistency when a client obtains the lock from one replica while the other client obtains that lock from another replica, cf. Appendix Section~\ref{sec:cvf-more}.) However, overlapping lock intervals do not necessarily mean the two clients accessed the shared data (protected by the lock) simultaneously because a client might want to obtain several locks before it started accessing the data. Furthermore, even if the clients accessed the data simultaneously, that does not necessarily mean the computed results would be wrong (the colors of neighboring nodes might still be different).

We validated this hypothesis with experiments where we ran 
the \coloring program for social graphs in \rollback mode.
We also added instrumentation to record information about the \cvf{s} such as the time intervals when the clients accessed shared variables and the colors of graph nodes computed by the clients. We analyzed the recorded data after the experiments had finished and found that among 116 \cvf{s} detected in the experiments, the client access intervals did not overlap in 35 of them. In the 81 \cvf{s} where the clients could have accessed shared variables simultaneously, only in 6 \cvf{s} that the computed colors were conflicting. (The reason the colors of two conflicting neighbors were still different even two clients were updating them simultaneously is that the color of a node was influenced by the colors of all of its neighbors, not just only the neighbor where the access conflict occurred.) So in most of the \cvf{s} we observed, the conflicts were resolved favorably.
These results imply that in rollback approach the program was inherently required to rollback more often than necessary (each detected \cvf caused a rollback) whereas in stabilization approach the program only had to handle a few actual faulty \cvf{s}. We believe this is one of the reasons why stabilization performed better than rollback in our experiments.

We note that the overhead of the above analysis is expensive and currently not suitable to be used with runtime rollback. It is an open problem to find  efficient mechanisms to do it.

\color{black}

\subsection{Results with Non-Stabilizing Algorithm}
\label{subsec:result-non-stabilizing-algo}

A natural question could be that what options should we choose if both stabilization and non-stabilization algorithms are available? We note that in general, when the algorithms are different, it is hard to fairly compare the two approaches since the performance is also affected by other factors such as optimization and implementation techniques. However, if the algorithms are closely similar, the comparison is useful. In this paper, we also compare the stabilization and non-stabilization algorithms for graph coloring since the algorithms are fairly similar.
(Our non-stabilization graph coloring algorithm is based on \cite{Raynal2013distributed}.)


Table \ref{tab:compare-with-non-stabilizing-algo} shows experiment results when running those algorithms on a regular random graph with 50,000 nodes, using normal partitioning scheme, in our local lab network with 20 \si{\ms} latency. 
%
A key observation from this analysis is that the stabilizing algorithm is less efficient than the non-stabilizing counterpart on sequential consistency.
However, it is the overall winner when used with eventual consistency, as it can benefit from tolerating \cvf{s}. By contrast, non-stabilizing algorithm cannot benefit from 
tolerating \cvf{s} 
thereby resulting in lower performance even with rollback. For example, for $d=10$ ($d$ is the average node degree), time taken by the  non-stabilizing algorithm was 7,021 \si{\second} in sequential consistency and it improved to 5,456 \si{\second} with eventual consistency and rollback. 
By contrast, the cost of the stabilizing algorithm under sequential consistency was 11,146 \si{\second}. It improved to 1,717 \si{\second} under EVE-AS model. 

This implies that while there may be some cost associated with making the protocol stabilizing, it is recovered by tolerating \cvf{s}. In this context, we also want to remind the reader that non-stabilizing algorithms cannot ignore \cvf{s}, as a \cvf may perturb the program to a state from where recovery is not guaranteed. Only stabilizing programs can choose to ignore \cvf{s} as they are designed to recover from them. Non-stabilizing programs can only use the detect-rollback approach to deal with \cvf{s}. 



\begin{table}[htbp]
\caption{Computation time (in seconds) of Stabilizing and Non-Stabilizing algorithms for graph coloring. 
The average of node degree ($d$) varies between 2 and 10. The baseline for calculating benefit is \seq}
\begin{tabular}{|p{1.35cm}|p{1.85cm}|r|r|r|r|}
\hline
\multicolumn{2}{|c|}{\parbox{2.5cm}{Average node degree}} & $d$=2 & $d$=3 & $d$=6 & $d$=10 \\
\hline
\multirow{3}{*}{\parbox{1.35cm}{\centering{Stabilizing graph coloring}}} 
    & SEQ       & 2,325 & 3,378 & 6,518 & 11,146 \\ 
    & Rollback  & 1,559 & 2,279 & 4,742.3 & 10,150 \\  
    & EVE-AS    & 1,321 & 1,376 & 1,615 & 1,717 \\ 
    \cline{2-6}
    & EVE-AS benefit    & $\times 1.8$ & $\times 2.5$ & $\times 4.0$ & $\times 6.5$ \\ 
\hline
\multirow{2}{*}{\parbox{1.35cm}{\centering{Non-stab. graph coloring}}} 
    & SEQ       & 1,653 & 2,291 & 4,246 & 7,021 \\
    & Rollback  & 1,213 & 1,681 & 3,192 & 5,456 \\
    \cline{2-6}
    & Rollback benefit & 26.6\% & 26.6\% & 24.8\% & 22.3\% \\
\hline
\end{tabular}
\label{tab:compare-with-non-stabilizing-algo}
\end{table}

\section{Related Work}
\label{sec:related-work}






\textbf{Distributed graph computation and consistency}. The availability of large-scale real-world graphs (social, biological, collaborative, etc.) facilitates the development of various graph computation engines~\cite{MWM.thinkvertex.2015ACMSurvey}.
These frameworks rely on sequential consistency to guarantee result correctness~\cite{GraphLab2012VLDB}. However, due to CAP theorem~\cite{CAP.Brewer00PODC,CAP.GL02SIGACT}, the performance of a fault-tolerating application will suffer if sequential consistency is maintained (given that fault tolerance is a must). Weaker consistency models such as causal, FIFO, etc.~\cite{Ghosh14} improve performance by relaxing the consistency requirement. Among them, eventual consistency~\cite{Vogels2009CACM} only guarantees that the replicas will eventually convergence once faults stop, and let the applications choose how to resolve data anomalies.



\textbf{Predicate detection and rollback.} The overall predicate detection framework is presented in~\cite{MN91WDAG}. Although the problem of predicate detection is NP-hard in general, efficient detection algorithms exist for some classes of predicates~\cite{CG98DC,CGNM13SRDS,mb15ipdps}. Once a violation is detected, the application is recovered to a previous state. Besides the typical snapshot-based rollback~\cite{DCDFC14OSDI,CADK17ICDCS}, it is possible to undo a violation by rolling back only those clients involved in the conflicts~\cite{NCKD2019JBCS}.

\textbf{Stabilization}. The notion of stabilization appeared in Dijkstra's seminal paper~\cite{EDW426}, and consists of two properties: convergence and closure~\cite{AG1993TSE,Dolev2000}. Furthermore, most of them are silent stabilization~\cite{DGS1999AI}. 
Also, there exist variants of stabilization such as weak, probabilistic, active, fault-containment stabilization \cite{Gouda01SSS,Herman90IPL,GGHP1996PODC,Kohler2012DC,bk11sss}.

Our work contributes to the existing literature by a comparative study and analysis of the performance of stabilization and rollback in handling \cvf{s} -- data anomalies that occur when a stabilizing graph application runs on eventually consistent key-value store.

\color{black}

\section{Conclusion}
\label{sec:concl}

In this paper, we considered the passive node model introduced in \cite{NKD2019ICDCN} and two approaches to reduce the time for execution of graph algorithms in it.
Since the use of eventual consistency has the potential to reduce execution time, we focused on managing (rather than eliminating) the inconsistency (denoted by consistency violating faults (\cvf{s}) in this model).
The first approach relied on detecting \cvf{s} and rolling back. The second, applicable only to stabilizing programs, was to observe that \cvf{s} are a subset of transient faults and, hence, are already tolerated {although at the cost of increased computation time for convergence}. 
Our analysis shows that for stabilizing programs, the second approach provides substantial benefits compared with the first one. Specifically, the second approach provides a 25\%--35\% improvement for different programs. Furthermore, the aggressive stabilization (that introduces additional \cvf{s} at the cost of efficiency) reduces the convergence time 2--15 times.
By contrast, the rollback based approach provides limited benefits and potentially causes performance to suffer when compared with sequential consistency. 

We also considered another approach to reduce the time for execution. It relied on heuristics to allow clients to keep track of nodes that may have enabled actions. Experimental results show that the heuristics can improve convergence time about 44\%
by reducing tail latency where the state of a very few nodes is inconsistent.

We also find that the stabilization based approach can benefit even more if the program can use other techniques to reduce overall time. Specifically, we considered the use of graph partitioning to reduce \cvf{s}. In this case, both approaches showed benefits but the benefit of 
stabilization was higher.

Another key insight in this work is that the benefits that apply for stabilizing algorithms can make them attractive in eventually consistent data stores even if they are (relatively) inefficient under sequential consistency. For example, in Section \ref{subsec:result-non-stabilizing-algo}, we showed that under sequential consistency, the stabilizing program had 58\% lower performance than a similar non-stabilizing program (11,146 \si{\second} to 7,072 \si{\second}). However, its performance was 3.2 times better under eventual consistency (1,717 \si{\second} vs 5,456 \si{\second}). This happened because the non-stabilizing algorithm could not tolerate \cvf{s} in the same manner that a stabilizing program could. This indicates that there may be a substantial benefit in 
{revising an existing algorithm for the problem at hand to make it stabilizing and}
reduce the overall runtime under eventual consistency.
{We note that there are several algorithms to \textit{add} stabilization to a non-stabilizing program~\cite{KP93DC}. These could be used in this context. However, an approach that optimizes the addition of stabilization using specific insight into the problem at hand may be more desirable as it is likely to provide the most benefit.}


As another demonstration,
consider the task of analyzing large-scale real-life networks (e.g. social networks) which are 
challenging to deal with. 
One of the challenges is that their complex structure imposes a significant locking coordination overhead for atomicity assurance, which retards the overall performance.
Most of existing work tried to reduce this overhead by efficient partitioning schemes~\cite{MWM.thinkvertex.2015ACMSurvey} but the improvement was limited due to the inherent complex graph structure and required a preprocessing step. In this paper, we observed that aggressive stabilization (\eveas) performed particularly well in social graphs (an order of magnitude improvement) without additional preprocessing overhead. This observation suggests that eventual consistency and stabilization is a promising candidate to efficiently tackle the complexity in social networks.


From the analysis of this work, we find that stabilization-based approach provides a substantial benefit compared with rollback-based approach. However, in both cases, the time required for convergence of the last few nodes is still quite high. One of the future work in this area is to reduce this overhead.  
Another future work is to generalize the results in this paper specifically to determine which options one should choose if both stabilizing and non-stabilizing algorithms are available.
{Another question for investigation is whether the analysis holds for other models of distributed computation.}


\section*{Acknowledgment}

We are grateful to late Professor Ajoy K. Datta for discussion that lead to development of ideas in this paper. 


\bibliographystyle{IEEEtran}
\bibliography{DuongNguyen}

\newpage
\appendix

\section{Appendix}
\label{sec:appendix}

In this Appendix, we provide a more detailed description of some issues that we are only able to briefly discuss in the main paper due to space constraints. Specifically, in Section~\ref{sec:voldemort-operation}, we describe the operation of Voldemort key-value store. Section~\ref{sec:sys-config-more} details the machine configuration and proxy implementation for the experiments. Next, we describe how 
a client executes a node action and how locks are used to avoid simultaneous updates by clients.
Then we explain the notion of \cvf both at abstract level as well as in concrete examples in Section~\ref{sec:cvf-more}. The termination detection algorithm used to measure convergence time of case study programs is described in Section~\ref{sec:termination-detection-algo}. We present the heuristics for improving the convergence time of stabilizing programs in Section~\ref{sec:result-extra-var}. Finally, Section~\ref{sec:compare-normal-random-partition} compares the normal partitioning scheme with its random counterpart.

\subsection{Voldemort Key-Value Store Operation}
\label{sec:voldemort-operation}

Voldemort is an open-source implementation of Amazon Dynamo key-value store~\cite{Dynamo.DHJKLPSVV07SOSP}. In this section, we describe the operation of Voldemort.

Every data entry in Voldemort is stored as a pair $<k,v>$ where $k$ is the key (unique name) and $v$ is the corresponding value of the key $k$. When using Voldemort to implement the \pasnode model for graph computation discussed in Sections~\ref{sec:intro} and~\ref{sec:sys-model-architecture}, the key is a node identifier and the value is the values of variables associated with that node (we concatenate the variable values as a single string delimited by separators).
Every client access to Voldemort key-value store is performed through two operations: the GET($k$) operation returns the current value of key $k$, and the PUT($k, v$) operation updates the key $k$ with new value $v$.

When a client wants to execute an action of the form $g \longrightarrow st$, it identifies all the variables required to execute this action. It issues a GET (i.e. read) request to all replicas (denoted by $N$, henceforth). It waits for receiving replies from at least $R$ --a configurable parameter in Voldemort-- replicas. If at least $R$ replicas reply before the timeout, the GET request is considered successful. If not, the client issues a second round of GET requests to the replicas. After the second, if replies are received from at least $R$ replicas in total (including the first round), the GET request is successful. Otherwise, it is not successful. If all reads are successful and the guard evaluates to true, the client identifies all variables that need to be changed. It then issues a PUT (i.e. write) request to all $N$ replicas. Similar to GET request, a PUT request is considered successful only if the client receives replies from at least $W$ --another configurable parameter in Voldemort-- replicas before timeout after at most two rounds. When the write is successful, action execution is complete. In the \pasnode model, the client does not have to retry an unsuccessful action. 

The clients learn the parameters $N$, $R$, and $W$ from the replicas at the time of startup. The above replication scheme employed by Voldemort is the active replication where the clients are in charge of data replication. The clients can also tune the values of $N$, $R$, $W$ if needed. By adjusting the value of $W$, $R$, and $N$, the consistency model of the key-value store is changed. For example, if $W + R > N$ and $W > \frac{N}{2}$ for every client of the same program, then the program is running on sequential consistency. If $W + R \le N$ then it is eventual consistency.
%
%

\subsection{System Configurations in the Experiments}
\label{sec:sys-config-more}

The local lab computer system consisted of 9 commodity PCs whose hardware configurations are specified in Table \ref{table:machine-config} and the machines were in the same local network. 
Three of the PCs were dedicated to the Voldemort servers (replicas) and six other PCs were shared by the clients (each client machine hosted multiple Voldemort client programs). The number of servers was 3 and the number of clients was 30. With three servers, we chose N3R1W1 (the number of replicas N=3, the number of required reads R = 1 and the number of required writes W = 1) for eventual consistency, and N3R1W3 for sequential consistency since in our experiments N3R1W3 had better performance than N3R2W2. Hereafter, for brevity, we use R1W1 and R1W3 instead of N3R1W1 and N3R1W3 respectively.

We deployed the experiments on our local lab since we could control some parameters such as network latency between the clients and servers. To adjust the network latency, we place a proxy process within each client machine. The proxy will relay all communication between the clients and the servers.
The proxy for client $C$ runs in parallel with client $C$ on the same machine.
When $C$ sends a message to server $S$, the message is routed through the proxy.
The proxy buffers the message for the required delay before forwarding it to $S$.
Responses from $S$ to $C$ is also handled by the proxy in the same manner.
This allows us to evaluate the protocols in different network delay scenarios. 
A more detailed description of proxy implementation is provided in \cite{NCKD2019JBCS}.

Besides the experiments on the local lab network, we also ran experiments on Amazon Web Services (AWS) network to confirm the results in a more realistic environment. In AWS experiments, we used three EC2 M5.xlarge instances for the servers and six EC2 M5.large instances for the clients (cf. Table \ref{table:machine-config}). 
The AWS machines are distributed in three regions (clusters): US East Ohio, US West Oregon, and Canada Central. The one-way latency among the AWS regions in our experiments (measured using \texttt{ping} command) were:
US West Oregon and US East Ohio: 26 ms
US West Oregon and Canada Central: 32 ms
Canada Central and US East Ohio: 15 ms
The average latency between AWS regions is about 24 ms.

\begin{table}[ht]
\caption{Configurations of machines used in the experiments}
\vspace{-15pt}
\begin{center}
\begin{tabular}{|p{1.3cm}|p{1.5cm}|p{2.3cm}|l|l|}
    \hline
    Environment & Machine & CPU & RAM & Storage \\
    \hline
    \multirow{3}{*}{Local lab} & 3 server machines & 8 Intel Core i7-4770T 2.50 GHz  & 8 GB & SSD \\
    \cline{2-5}
    & 5 client machines & 4 Intel Core i5 660 3.33 GHz & 4 GB & HDD \\
    \cline{2-5}
    & 1 client machine  & 4 Intel Core i5-2500T 2.30GHz & 4 GB & HDD \\
    \hline
    \multirow{2}{*}{AWS} & 3 server machines (EC2 M5.xlarge) & 4 vCPUs  & 16 GB & SSD\\
    \cline{2-5}
    & 6 client machines (EC2 M5.large) & 2 vCPUs & 6 GB & SSD \\
    \hline
\end{tabular}
\label{table:machine-config}
\end{center}
\end{table}

\subsection{Executing a Node Action by Client}
\label{sec:graph-locks}

The procedure for a client $C$ to process a node $i$ assigned to its partition is as follows:
\begin{itemize}
    \item[(1)] Obtain exclusive update privilege for the state of node $i$ and read privilege for neighbors of $i$ (i.e. no other client should read the state of $i$ or update the state of $i$ neighbors).
    \item[(2)] Read the state of $i$ (variables of $i$) and its neighbors.
    \item[(3)] Compute the new values for $i$'s variables.
    \item[(4)] Write the new state of $i$ to the store (this step can be omitted if all of $i$'s variables are unchanged), and 
    \item[(5)] Release the privileges it holds for $i$ and its neighbors.

\end{itemize}

In order to support client $C$ in obtaining necessary privileges, we designate one Peterson lock~\cite{Peterson81IPL} associated with each graph edge.
To read the state of node $i$, client $C$ just needs to obtain a lock associated with any edge incident on $i$.
To update the state of node $i$, however, client $C$ needs to obtain the locks associated with all edges incident on $i$. Once such locks are obtained by $C$, other clients can read state of $i$'s neighbors but they cannot update any of them (since they have to wait for one of the locks being hold by $C$).

\begin{figure}[h]
    \centering
    \includegraphics[width=0.36\textwidth]{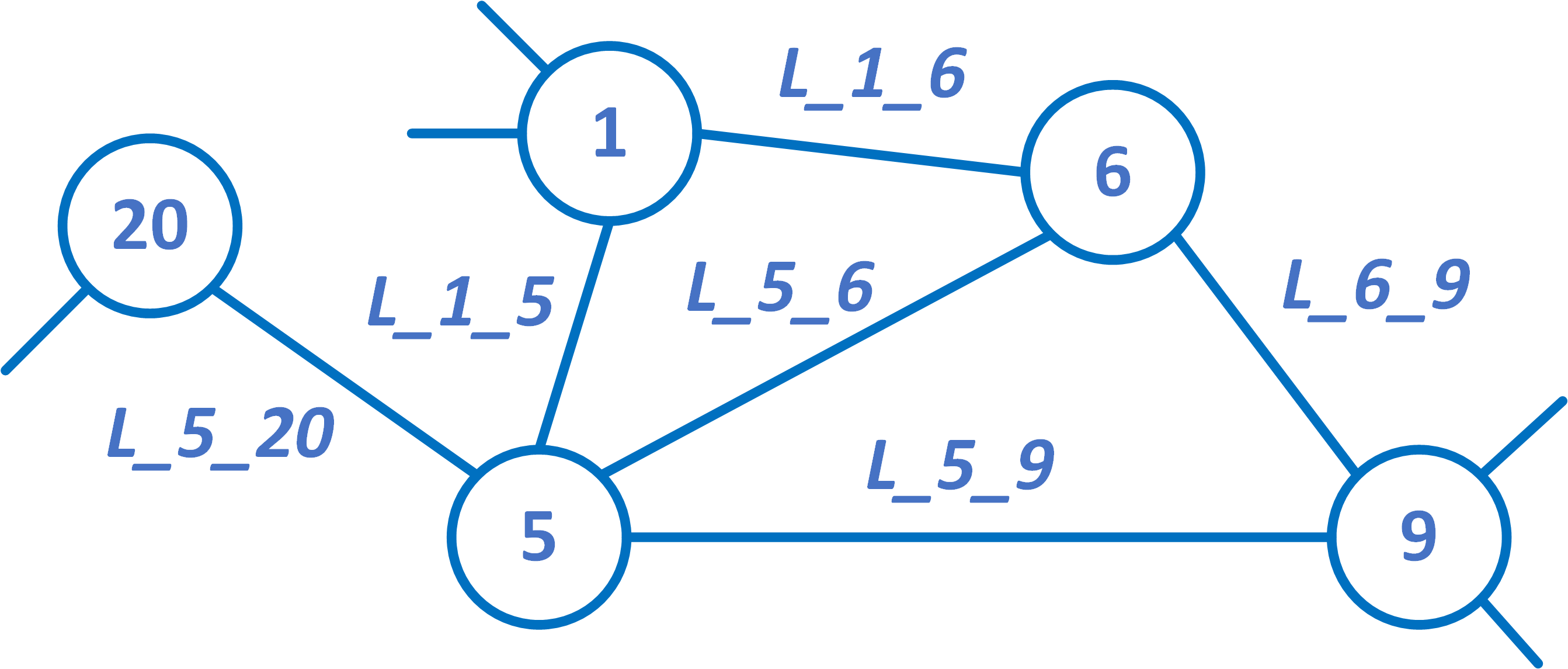}
    \caption{Illustration of locks. To update node 6, a client has to obtain these locks in following order: $L\_1\_6, L\_5\_6, L\_6\_9$}
    \label{fig:lock-example}
\end{figure}

For deadlock avoidance, client $C$ obtains the required locks in lexicographical order. Suppose $i < j$ then the lock for edge $(i,j)$ is $L\_i\_j$. As an illustration, in Figure ~\ref{fig:lock-example}, if client $C$ wants to update node $6$, it has to obtain these locks in the following order $L\_1\_6, L\_5\_6, L\_6\_9$.

Once $C$ has had the update privilege for node $6$, no other client can update any neighbor of $6$ (says $5$) since that client will have to wait for $C$ to release the lock $L\_5\_6$.

We note that the above locking scheme only works if the shared data is sequentially consistent. If it is eventually consistent, simultaneous updates could happen as explained in Section~\ref{sec:cvf-more}.

Since obtaining locks constitutes a sizeable proportion of client computation time (waiting for other clients to release the required locks) and many nodes shared neighbors, a client usually does not process each node individually but processes a batch of nodes at the same time.
This batch processing reduces the number of locks the client needs to obtain. For example, suppose nodes 5 and 6 are assigned to a client. If the client processes 5 and 6 individually, it will have to obtain 5 locks, namely $L\_1\_5, L\_5\_9, L\_5\_20$ (for node 5) and $L\_1\_6, L\_6\_9$ (for node 6). Note that there is no need for obtaining lock $L\_5\_6$ if both nodes are assigned to the same client. On the other hand,
if the client processes node 5 and 6 together, it only has to obtain 3 locks, namely either $L\_1\_6$ or $L\_1\_6$, $L\_5\_20$, and either $L\_5\_9$ or $L\_6\_9$. 

\color{black}

\subsection{Consistency Violating Faults (\cvf)}
\label{sec:cvf-more}

In the \pasnode model, the program state is stored at the replicas. The protocol for synchronizing replicas can be passive replication or active replication (the case of Voldemort). 


In passive-replication-based sequential consistency, the protocol enforces that all replicas are strictly synchronized. A replica will not provide the new value unless that value has been committed by other replicas. A client reading from any of the replicas will always obtain the fresh data. However, for eventual consistency, the protocol is relaxed and allows replicas to return the current values they know, which may be not up-to-date. 

In active-replication-based sequential consistency, the protocol requires each update to be committed by a majority of replicas and the client reads from at least one of them, thus obtains the fresh data. For eventual consistency, however, the protocol is relaxed where the read and write quorums do not overlap. Thus, some replicas may have not received the latest updates due to transient faults, and if a client reads from those replicas, it obtains a stale value.

In short, a client always obtains the fresh data with sequential consistency and may obtain a stale data with eventual consistency.
Reading stale information could lead the clients to incorrect computation steps/transitions.

\textbf{Abstract description of \cvf. } For each variable $x$ of node $j$, each replica $i$ maintains a value of $x.j$ as a key-value pair. For the purpose of illustration, assume that there are three replicas and the values of $x.j$ at these replicas are $r_1, r_2$ and  $r_3$. Denote $f(r_1, r_2, r_3)$ as the abstract value of $x.j$ where $f$ is some resolution function that chooses a value among $r_1, r_2, r_3$ in a deterministic manner.
For example, function $f$ chooses the latest value of $x.j$ (assume that each value is also associated with a logical or physical timestamp).
%
In sequential consistency where the replication protocol provides the impression that all the replicas work as if there is only a single replica, access (read/write) to variable $x.j$ by any client always returns the same abstract value of $x.j$. In eventual consistency, however, this property may be violated when different clients observe different values of $x.j$ (e.g. client $c1$ observes value $r_1$ while client $c2$ observes value $r_2$). Only one of those values is up-to-date and the other is stale.
We also note that reading stale data is possible in eventual consistency but such anomalies are expected to be not frequent~\cite{Dynamo.DHJKLPSVV07SOSP} and they are usually associated with transient faults.



\begin{figure}
    \centering
    \includegraphics[width=0.46\textwidth]{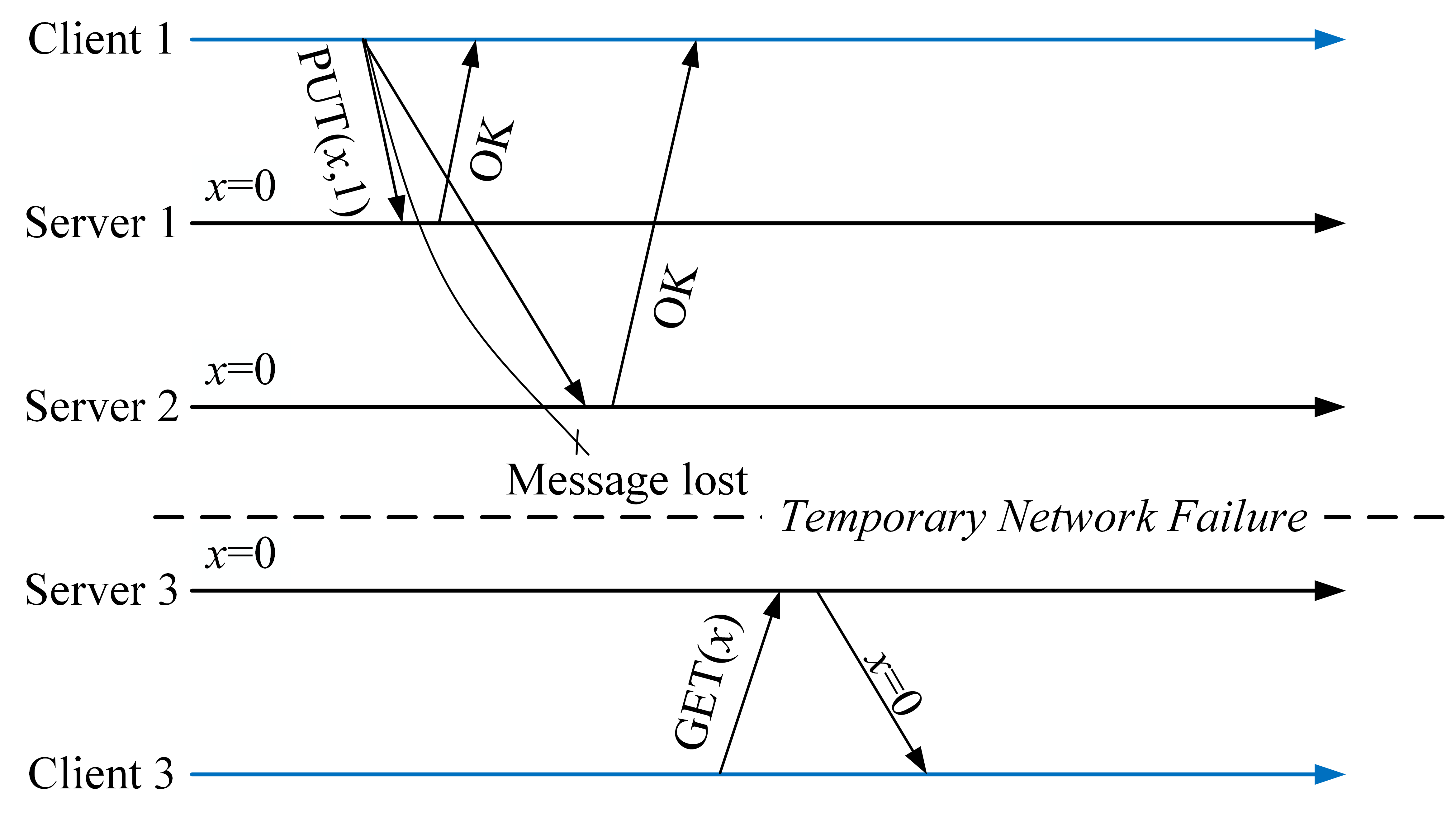}
    \caption{Illustration of \cvf in Voldemort. Clients run on eventual consistency R1W1}
    \label{fig:how-cvf-occurs}
\end{figure}

\textbf{Reading stale values due to eventual consistency in Voldemort. } Figure~\ref{fig:how-cvf-occurs} illustrates how a \cvf occurs in Voldemort key-value store where the clients are running on eventual consistency R1W1. Suppose $x=0$ initially. Client 1 updates the value of $x$ to 1 by sending PUT($x,1$) request to all replicas/servers. Due to a temporary network failure, the request does not reach server 3. However, the PUT request still succeeds since client 1 receives replies from two servers (PUT request is successful if client receives at least W=1 replies). If client 3 reads the value of $x$, it will obtain the stale value $x=0$ from server 3 (this read succeeds since client 3 just needs R=1 response). In contrast, any client served by server 1 and server 2 will see the new value $x=1$.

Reading stale values can lead to erroneous transitions.
For example, in the arbitrary graph coloring problem (\coloring) suppose client 1 wants to work on node 5 while client 3 wants to work on node 20 (cf. Figure~\ref{fig:lock-example}). If variable $x$ in Figure~\ref{fig:how-cvf-occurs} is the shared lock $L\_5\_20$, then client 1 will think that it has obtained the lock while client 3 observes the lock is still vacant and tries to obtain it (it will succeed since one confirmation from server 3 is sufficient). As a result, both clients enter a critical section simultaneously.
Suppose the initial color of each node is color 0 and the two clients read these values. Then the clients will likely update the colors of both node 20 and node 5 to color 1, resulting in a new invalid coloring.
We note that if locks are not used to guarantee atomicity (such as in \eveas mode), client 1 and client 3 may also read and update the color of node 5 and node 20 simultaneously without knowing so, and produce invalid coloring results in a similar manner.


\textbf{Computation in the presence of \cvf. } With the introduction of \cvf, the computation of program $p$ in a eventually consistent replicated \pasnode model is of the form \br{s_0, s_1, \cdots} where

\begin{itemize}
\item  $\forall l: l \geq 0: $, $s_l$ is a state of $p$,
\item $\forall l: l \geq 0: (s_l, s_{l+1}) \in \delta_p \cup \cvf_p$ or\\ $(s_l = s_{l+1})$ and no action of $p$ is enabled in state $s_l$, and
\item If some action $ac$ of $p$ (of the form $g \longrightarrow st$) is continuously enabled (i.e., there exists $l$ such that $g$ is true in every state in the sequence after $s_l$) then $ac$ is eventually executed (i.e., for some $x \geq l$, $(s_x, s_{x+1})$ corresponds to execution of $st$.) 

\end{itemize}



\subsection{Termination Detection Algorithms.} 
\label{sec:termination-detection-algo}

Our termination detection algorithm to determine whether a program has reached a fixed point in the computation is based on the algorithm in \cite{DijkstraFG1983IPL}. We briefly describe the termination detection algorithm. Basically, the termination detector is also a Voldemort client program running a detection algorithm consisting of two rounds. In the first round, the algorithm reads the state of all nodes (including modification timestamps) and determines if every node has become disabled (i.e., all of its actions have the guards be evaluated to false). If that is true, it moves to the second round; otherwise, it restarts the first round.
In the second round, the algorithm checks if the state and modification timestamp of every node is unchanged since the most recent successful first-round check. If there is any change, the algorithm restarts from the first round; otherwise, it reports the termination of computation. 
The termination detector runs in the consistency mode where $R = N$ (the number of required reads equals the number of replicas) to ensure reliability.
Since the termination detector only reads from and does not write to the key-value store, it minimally affects the stabilization time of the computation.

\subsection{Improving the convergence time in stabilization approach}
\label{sec:result-extra-var}

\textbf{Heuristics to reduce tail latency. }
In the passive node model, clients are responsible for checking which nodes have enabled actions and execute those actions. The results in Table \ref{tab:self-stabilization-vs-rollback} correspond to the case where clients evaluated the guards of nodes assigned to them in a round-robin manner. One of the issues with round-robin is that some nodes whose actions are enabled may not be considered while the client is evaluating other nodes assigned to it but having no enabled actions. Note that this issue is ignored in the active node model, as, generally, it is assumed that the scheduler will choose some active node for execution. The time required for the scheduler to determine this node is ignored. 

In the programs under consideration, if no action of $j$ is enabled in the current state then this information is stable until a neighbor of $j$ executes. Thus, if a node could tell the client that its actions are unlikely to be enabled then the client can save on reading the states of its neighbors. For such an approach to work, for node $j$, we need to know (1) $nd\_change.j$  the last time the client checked that no actions are enabled at $j$, and (2) $nbr\_change.j$ the last time one of its neighbors was updated. 

Thus, when client reads the state of $j$ and finds that $nd\_change.j > nbr\_change.j$, it does not need to read the state of its neighbors to determine if some action of $j$ is enabled. Since clocks of all computers involved may not be identical, we change the condition to $nd\_change.j > nbr\_change.j + \Delta.j + \epsilon$ where $\Delta.j$ is the length of the last execution of $j$ and $\epsilon$ is the upper bound for clock synchronization error. In other words, if $nd\_change.j > nbr\_change.j + \Delta.j + \epsilon$  is true then the client can save time by not issuing GET requests to neighbors of $j$. 

\begin{figure}[!t]
    \centering
    \includegraphics[width=0.46\textwidth]{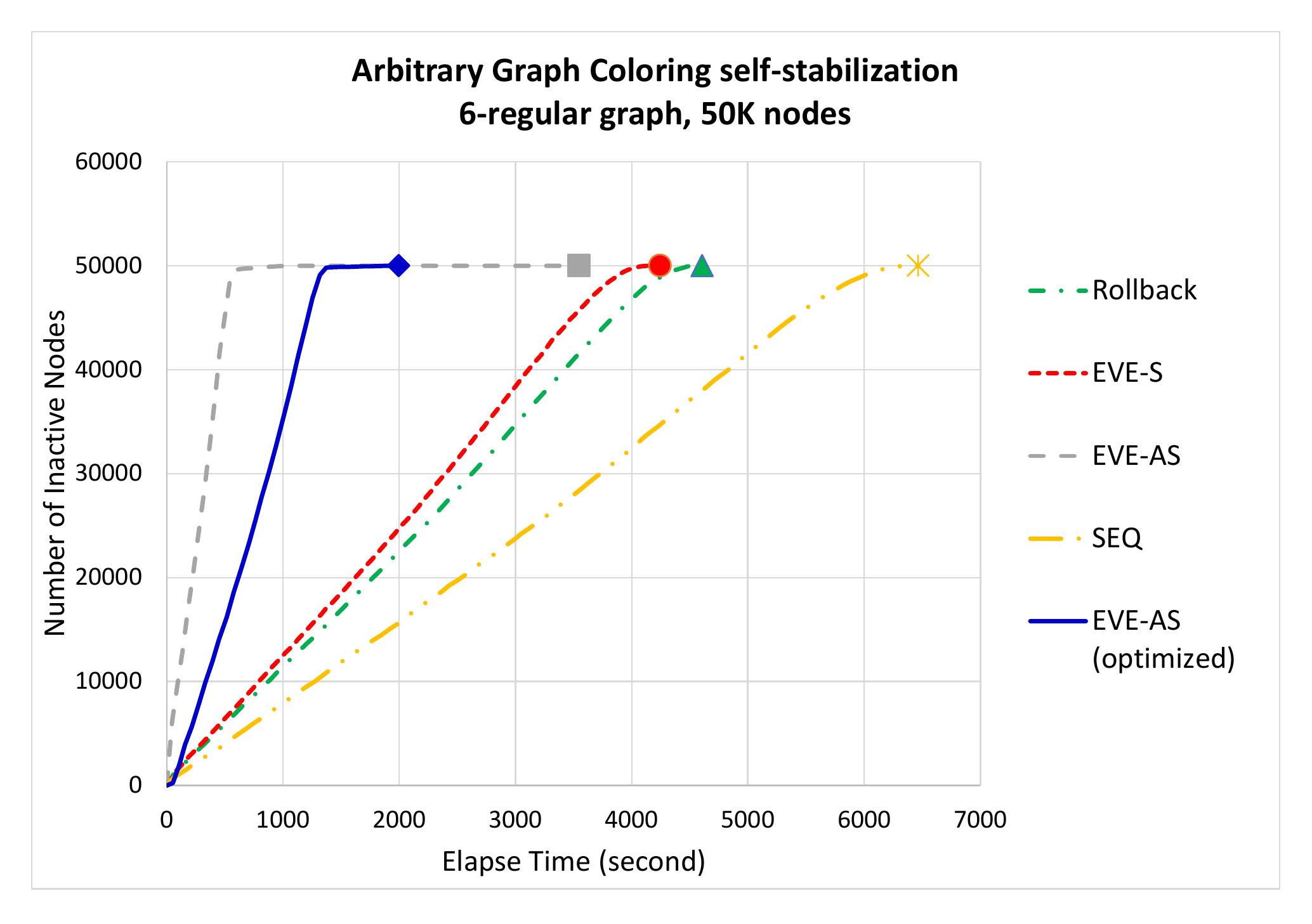}
    \caption{The convergence pattern of different execution modes in \coloring. Normal partitioning. Latency was 20 \textit{ms}.}
    \label{fig:arbitrary-coloring-convergence-graph}
\end{figure}

Table \ref{tab:stabilization-optimization} considers execution with this optimization. We find that this optimization is useful only when the convergence pattern exhibits a long tail at the end (cf. EVE-AS mode in Figure \ref{fig:arbitrary-coloring-convergence-graph}). 
The overhead of the optimization (for reading and writing additional variables) caused EVE-AS (optimized) to converge slower than EVE-AS at first. However, the optimization significantly reduced the tail of convergence graph and thus improved the overall convergence time by 44\%. If the convergence pattern did not have the long tail characteristic (such as EVE-S mode in Figure \ref{fig:arbitrary-coloring-convergence-graph}, or EVE-AS mode with random coloring), this optimization increased the convergence time because of the extra overhead (cf. Table~\ref{tab:stabilization-optimization}).

\begin{table}[!t]
\centering
\caption{Effectiveness of the random coloring and the optimization for stabilization approach in the arbitrary graph coloring problem (\coloring). Convergence time is measured in seconds. Normal partition. Latency = 20 \textit{ms}}
\vspace{-5pt}
\begin{tabular}{|p{1cm}|p{1.4cm}|p{1.4cm}|r|r|} 
\hline
Execution mode & \multicolumn{1}{c|}{\parbox{1.5cm}{Track update timestamp}} & \multicolumn{1}{c|}{\parbox{1.4cm}{Color selection scheme}} & \multicolumn{1}{c|}{\parbox{1.3cm}{Regular graph 50K}} & \multicolumn{1}{c|}{\parbox{1.3cm}{Social graph 50K}} \\
\hline
EVE-AS & Yes   & Deterministic & 1,972 & 4,805 \\
EVE-AS & Yes   & Random        & 1,941 & 4,807 \\
EVE-AS & No  & Deterministic & 3,547 & 1,885 \\
EVE-AS & No  & Random        & 1,431 & 1,883 \\
\hline
EVE-S  & No   & Deterministic & 4,270 & 18,229 \\ 
EVE-S  & Yes   & Deterministic & 5,136 & \textgreater 20,000 \\
\hline
\end{tabular}
\label{tab:stabilization-optimization}
\end{table}

\textbf{Randomization. } As discussed in Section~\ref{sec:result-basic:ss-vs-rollback}, we observed some \cvf{s} when running \coloring in aggressive stabilization mode on regular graphs that prevented the program to converge.
For example, suppose two clients $C_1$ and $C_2$ are working on two nodes $v_1$ and $v_2$ at the same time. Suppose the original color of both nodes is 0. Because no mutual exclusion is used in \eveas, both clients may assign the same new colors 1 for both nodes, resulting in invalid/inconsistent coloring. 
This error is usually resolved when one of the clients visits its node in the next round and change its node to a different color. However, if both $C_1$ and $C_2$ re-visit $v_1$ and $v_2$ at the same time, the problem persists.
We observed this problem occurred only in regular graphs where the workload was split very evenly among the clients and there were only a few nodes with inconsistent colors that needed to be fixed. The problem did not happen in social graphs since the client workload was not even.
In other words, running \coloring in \eveas mode does not guarantee convergence. 

One possibility to address this problem is to modify the coloring algorithm so that the client would choose a random value among available colors for its nodes. With this modification, \eveas is probabilistic self-stabilizing. In our experiments with the random coloring scheme, the convergence time of coloring the same regular graph in \eveas improved from \SI{3,547}{seconds} to \SI{1,431}{seconds}. On the other hand, the convergence time for social graph stayed almost the same (1,885 \si{\ms} and 1,883 \si{\ms}, cf. Table~\ref{tab:stabilization-optimization}). 

We also note that the performance of sequential mode (\seq) is unaffected by whether deterministic coloring or random coloring is used (6,518 \si{\ms} and 6,544 \si{\ms}, not shown in Table~\ref{tab:stabilization-optimization}). Thus, for the \coloring program on random regular graphs, random coloring improves the benefits of \eveas when compared to deterministic coloring ($\times 4.6$ speedup vs. $\times 1.8$ speedup).

\subsection{Comparing Normal and Random Partitioning}
\label{sec:compare-normal-random-partition}

In Section~\ref{sec:result-basic:ss-vs-rollback}, we mentioned the benefits of random partitioning in evenly distributing the workload among clients. To quantitatively justify this argument, we partitioned a planar graph using normal and random partitioning schemes, and measured some properties of the resultants partitions (cf. Table~\ref{tab:planar-normal-vs-random-partition}).

\begin{table}[htbp]
\caption{Comparison between normal partitioning scheme and random partitioning scheme of a planar graph. For each property, the average (AVG) and standard deviation (STDEV) among the partitions are calculated.}
\vspace{-2pt}
\begin{tabular}{|p{2cm}|r|r|r|r|}
\hline
\multirow{2}{*}{\parbox{2cm}{Properties}} & \multicolumn{2}{c|}{Normal partition} & \multicolumn{2}{c|}{Random partition} \\
\cline{2-5}
&  AVG & STDEV & AVG & STDEV \\
\hline
Max degree      &  15.3   &5.0    &17.1   &2.7 \\
Min degree      &  2.7    &0.6    &1.0    &0.0 \\
Total degree  &  1622.2 &508.5  &1622.2 &59.9 \\
Node count   &  367.8  &4.7    &367.8  &4.7 \\
Average degree     &4.4    &1.4    &4.4    &0.1 \\
External edges &585.7  &148.9  &1568.1 &54.4 \\
Internal edges &1036.5 &508.7  &54.1   &13.6 \\
\hline
\end{tabular}
\label{tab:planar-normal-vs-random-partition}
\end{table}

We observed that properties related to nodes' degrees were more evenly distributed with random partitioning. Since a node's degree reflects the cost of obtaining locks and reading state of neighbors, which constitutes a significant chunk of work, an even degree distribution implies more balanced workload among partitions/clients. Consequently, 
we avoided slow clients that were assigned too much work.

On the other hand, random partitioning potentially breaks the locality characteristics of planar graphs. We observed with random partitioning, the number of external edges that crossed between partitions increased whereas the number of internal edges that connected nodes within a partition decreased. This implies random partitioning increased the locking overhead. Consequently, computation time often increased with random partitioning (cf. Table~\ref{tab:self-stabilization-vs-rollback-planar-random-partition}).





\end{document}